\documentclass[a4paper]{article}
\usepackage[german, english]{babel}
\usepackage{a4wide}
\usepackage{amsmath}
\usepackage{amssymb}
\usepackage{ifthen}
\usepackage{epsfig}
\usepackage[hang,nooneline]{subfigure}
\newcounter{fig}

\begin{document}

\title{\bf Compact Boson Stars}
\vspace{1.5truecm}
\author{
{\bf Betti Hartmann$^a$, Burkhard Kleihaus$^b$, 
Jutta Kunz$^b$ and Isabell Schaffer$^a$}\\
$^a$School of Engineering and Science, Jacobs University, Postfach 750 561\\
D-28725 Bremen\\
$^b$Institut f\"ur  Physik, Universit\"at Oldenburg, Postfach 2503\\
D-26111 Oldenburg, Germany\\
}

\vspace{1.5truecm}

\date{\today}

\maketitle
\vspace{1.0truecm}

\begin{abstract}
We consider compact boson stars that arise for a V-shaped 
scalar field potential.
They represent a one parameter family of solutions
of the scaled Einstein-signum-Gordon equations.
We analyze the physical properties of these solutions
and determine their domain of existence.
Along their physically relevant branch emerging from the
compact $Q$-ball solution, 
their mass increases with increasing radius.
Empoying arguments from catastrophe theory
we argue that this branch is stable,
until the maximal value of the mass is reached.
There the mass and size are on the order of magnitude of the
Schwarzschild limit, and thus the
spiralling respectively oscillating behaviour,
well-known for compact stars, sets in.

\end{abstract}

\section{Introduction}

When a complex scalar field is coupled to gravity,
boson stars arise as stationary localized solutions
of the coupled Einstein-Klein-Gordon equations
\cite{Feinblum:1968,Kaup:1968zz,Ruffini:1969qy}.
The physical properties of the boson stars depend strongly
on the scalar field potential
(see e.g.~the review articles
\cite{Lee:1991ax,Jetzer:1991jr,Liddle:1993ha,Mielke:1997re,Schunck:2003kk}).

If only a mass term is present, for instance, but no self-interaction,
so-called mini boson stars arise.
Their masses are relatively small, being bounded
by a maximal mass $M_{\rm max}$ on the order of
the Planck mass squared, divided by the boson mass,
$M_{\rm Pl}^2/m_{\rm B}$
When a repulsive quartic self-interaction is included
larger boson stars are obtained,
having a maximal mass $M_{\rm max}$ on the order of
$\sqrt{\lambda} M_{\rm Pl}^3/m_{\rm B}^2$,
where $\lambda$ is the self-coupling constant
\cite{Colpi:1986ye}.
The presence of a sextic potential, on the other hand,
leads to boson stars with a maximal mass $M_{\rm max}$
on the order of $M_{\rm Pl}^4/m_{\rm B}^3$
\cite{Lee:1991ax}.
These solitonic boson stars even possess a flat space limit,
where they correspond to non-topological solitons \cite{Friedberg:1976me}
(or $Q$-balls \cite{Coleman:1985ki}).

All these types of boson stars, whether small or large,
do not possess a sharp radius.
The scalar field exhibits an exponential fall-off 
as dictated by the particle mass, but this does not yield a
unique boundary for the star.
Thus a large variety of radius definitions
are employed, when one is interested in extracting
the relation between the mass of the boson stars
and their size.

Here we consider boson stars, whose radius is uniquely
defined, because these solutions are compact.
Their scalar field vanishes identically outside
the radius of the star.
These compact boson stars arise when a V-shaped 
self-interaction potential is employed.
In the flat space limit 
this potential allows for compact $Q$-balls,
which represent solutions of the signum-Gordan equation for the scalar field
\cite{Arodz:2008jk,Arodz:2008nm}.
Interestingly, when the electromagnetic field is coupled,
even shell-like solutions appear,
where the scalar field vanishes identically outside a finite shell
\cite{Arodz:2008nm,Kleihaus:2009kr,Kleihaus:2010ep}.

By solving the coupled Einstein-signum-Gordon equations,
we determine the domain of existence of these compact boson stars.
In particular, we show that by scaling the equations appropriately,
the solutions depend only on a single parameter,
varying between zero and a finite maximal value.
We then analyze the physical properties of these
compact boson stars and address their stability
from a catastrophe theory point of view
\cite{Kusmartsev:2008py,Kusmartsev:1992,Tamaki:2010zz,Tamaki:2011zza,Kleihaus:2011sx}.

The paper is organized as follows.
In section 2 we present the action, the Ansatz,
the equations of motion together with the scaling property,
and the global charges.
We present our solutions in section 3,
and discuss their physical properties.
We end with a conclusion and an outlook
in section 4.

\section{Action}

We consider the action of a self-interacting complex scalar field
$\Phi$ coupled to Einstein gravity
\begin{equation}
S=\int \left[ \frac{c^4}{16\pi G}R
   -  \frac{\hbar^2}{m_0}\left( \partial_\mu \Phi \right)^* \left( \partial^\mu \Phi \right)
 - U( \left| \Phi \right|) 
 \right] \sqrt{-g} d^4x
 , \label{action}
\end{equation}
with curvature scalar $R$, Newton's constant $G$, 
a mass scale $m_0$,
and the asterisk denotes complex conjugation.
The scalar potential $U$ is chosen as
\begin{equation}
U(|\Phi|) =  2\hat{\lambda}  |\Phi| 
 . \label{U} \end{equation} 
The dimensions of $\Phi$ and $\hat{\lambda}$ are $1/length^{3/2}$,
respectively $[energy]/length^{3/2}$.
Next we introduce the scaled scalar field $\Psi = \hbar/\sqrt{m_0} \Phi$. 
The resulting action reads
\begin{equation}
S=\int \left[ \frac{c^4}{16\pi G}R
   -  \left(\partial_\mu \Psi \right)^* \left( \partial^\mu \Psi \right)
 - 2 \lambda \left| \Psi \right|) 
 \right] \sqrt{-g} d^4x
 , \label{action2}
\end{equation}
where we defined $\lambda= \sqrt{m_0}/\hbar \hat{\lambda}$.

Variation of the action with respect to the metric and the matter fields
leads, respectively, to the Einstein equations
\begin{equation}
G_{\mu\nu}= R_{\mu\nu}-\frac{1}{2}g_{\mu\nu}R = \frac{8\pi G}{c^4} T_{\mu\nu}
\  \label{ee} \end{equation}
with stress-energy tensor
\begin{eqnarray}
T_{\mu\nu} &=& g_{\mu\nu}{L}_M
-2 \frac{\partial {L}_M}{\partial g^{\mu\nu}}
\nonumber\\
&=& 
  - \frac{1}{2} g_{\mu\nu} \left((\partial_\alpha \Psi)^* (\partial_\beta \Psi)
  + (\partial_\beta \Psi)^* (\partial_\alpha \Psi)    \right) g^{\alpha\beta}
  + (\partial_\mu \Psi)^* (\partial_\nu \Psi) 
  + (\partial_\nu \Psi)^* (\partial_\mu \Psi)
\nonumber\\
& & 
    -2\lambda g_{\mu\nu}  |\Psi| 
 , \label{tmunu}
\end{eqnarray}
and the matter field equation,
\begin{eqnarray}
& &\nabla_\mu \nabla^\mu \Psi = - \lambda \frac{\Psi}{|\Psi|}
 , \label{feqH} \end{eqnarray}
where $\nabla_\mu$ denotes the covariant derivative.

To construct static spherically symmetric solutions
we employ Schwarz\-schild-like coordinates and adopt
the spherically symmetric metric
\begin{equation}
ds^2=g_{\mu\nu}dx^\mu dx^\nu=
  -A^2N c^2dt^2 + N^{-1} dr^2 + r^2 (d\theta^2 + \sin^2\theta d\phi^2)
 , \end{equation}
with
\begin{equation}
N=1-\frac{2G}{c^2}\frac{m(r)}{r}
 . \end{equation}

The Ansatz for the matter fields has the form
\begin{equation}
 \Psi = \psi(r) e^{i c\omega t}
 . \label{phi} \end{equation}

The resulting Einstein and field equations are
\begin{eqnarray}
\frac{dm}{dr} 
& = & 
\frac{4 \pi r^2}{c^2}
\left(N \left(\frac{d\psi}{dr}\right)^2
     + \frac{\omega^2}{A^2 N} \psi^2
     + 2 \lambda \psi \right)
\ , \label{dmdr}\\
\frac{dA}{dr} 
& = & 
\frac{8 \pi G r}{c^4}
\left(A \left(\frac{d\psi}{dr}\right)^2
     + \frac{\omega^2}{A N^2} \psi^2 \right)
\ , \label{dAdr}\\
\frac{d^2 \psi}{dr^2} 
& = & 
-\frac{\omega^2}{A^2 N^2}\psi +\frac{\lambda}{N} {\rm sign}(\psi)
-\frac{1}{A N r^2}\frac{d}{dr}\left(A N r^2\right) \frac{d\psi}{dr}
\ , \label{dPdrr}
\end{eqnarray}
In the next step we introduce the dimensionless coordinate $x=r/\omega$ and
the dimensionless scalar field $h= \omega^2/\lambda\psi$. This yields
\begin{eqnarray}
\mu'
& = & 
x^2\left(N h'^2
     + \frac{1}{A^2 N} h^2
     + 2 h \right)
\ , \label{mup}\\
A'
& = & 
2 \alpha x
\left(A h'^2
     + \frac{1}{A N^2} h^2 \right)
\ , \label{Ap}\\
h'' 
& = & 
-\frac{1}{A^2 N^2}h +\frac{1}{N} {\rm sign}(h)
-\frac{\left(A N x^2\right)'}{A N x^2} h'
\ , \label{hpp}
\end{eqnarray}

where 
\begin{eqnarray}
\alpha & = & \frac{4 \pi G}{c^4}\frac{\lambda^2}{\omega^4}\ ,
\\
\mu & = & \frac{\omega^5 c^2}{4 \pi \lambda^2} m =
          \frac{\omega G}{c^2 \alpha} m
\end{eqnarray}
are the dimensionless coupling parameter, respectively mass function and 
$N=1-2 \alpha \mu / x$. 

Let us now specify the boundary conditions for the
metric and matter functions.
For the metric function $A$ we adopt
\begin{equation}
A(x_0)=1
\ , \end{equation}
where $x_0$ is the outer radius, thus fixing the time coordinate.
For the metric function $N(x)$ we require
\begin{equation}
N(0)=1 \ , 
\end{equation}
and for the scalar function we require 
at the origin and at the outer radius $x_0$, respectively,
\begin{equation}
h'(0)=0 \ , \ \ \ h(x_0)=0 \ , \ \ \ h'(x_0)=0 \
. \end{equation}

Note that 
$h(x)=0, A(x)=1, \mu(x)=const$ is an exact solution of the
equations (\ref{mup})-(\ref{hpp}) in the exterior region $x \ge x_0$.

Hence the mass of the boson star is given by 
\begin{equation}
M = \frac{\alpha}{\omega}\frac{c^2}{G} \mu(\infty) 
  = \frac{\alpha}{\omega}\frac{c^2}{G} \mu(x_0) \ ,
\end{equation} 
and the areal radius of the boson star is 
\begin{equation}
R = x_0/\omega \ .
\end{equation} 
As a consequence, 
$\frac{2 G M}{c^2 R} = 2\alpha \frac{\mu(\infty)}{x_0}$.

The particle number $Q$ is computed from the density
\begin{equation}
\rho= \frac{i\hbar}{2 m_0} g^{tt} 
\left(\Phi^* \partial_t\Phi -\Phi\partial_t\Phi^*\right) \ .
\end{equation}
Using the ansatz and dimenensionless quantities yields
\begin{equation}
\rho=\frac{\lambda^2}{\omega^3}\frac{1}{\hbar c} \frac{h^2}{A^2 N} \ .
\end{equation}
Integration then gives the particle number
\begin{equation}
Q = \int\rho \sqrt{-g} dr d\theta d\phi
  = \frac{\lambda^2}{\omega^6}\frac{4\pi}{\hbar c} 
     \int_0^\infty \frac{h^2}{AN} x^2 dx
  = Q_0 \hat{Q} \ ,
\end{equation}
 where 
\begin{equation}   
  Q_0   = \frac{\lambda^2}{\omega^6}\frac{4\pi}{\hbar c} \ , \ \ \ \ 
  \hat{Q} = \int_0^\infty \frac{h^2}{AN} x^2 dx \ .
\end{equation}

For the central energy density we find
\begin{equation}
\epsilon_0 = \frac{\lambda^2}{\omega^2} \left(\frac{h_0^2}{A_0^2} + 2 h_0\right)
 = \frac{\lambda^2}{\omega^2} \epsilon_c \ ,
\end{equation}
where $h_0 = h(0)$, $A_0= A(0)$ and the boundary condition $h'(0)=0$ were
used.
The central mass density is then defined as $n_0=\epsilon_0/c^2$.

\section{Results}

We solve the coupled system of equations numerically
employing a Newton-Raphson scheme.

Let us begin our discussion of the results with
the flat space-time limit, the compact $Q$-ball.
Here the scalar field function can be obtained analytically
\cite{Arodz:2008jk,Arodz:2008nm}
\begin{equation}
 h= \left\{ 
\begin{array}{clr}
\displaystyle 1- \frac{x_0}{x} \frac{\sin x}{\sin x_0}
& {\rm if}\ & 0 \le x \le x_0 \\
0 & {\rm if}\ & x \ge x_0 \ ,
\end{array} \right.
\end{equation}
where $x_0$ satisfies $x_0=\tan x_0$, i.e., $x_0 \approx 4.4934$,
while the metric functions are trivial, $N(x)=A(x)=1$.
This analytical solution is our starting point,
obtained in the limit $\alpha \to 0$.
It is seen in Figs.~1a-1c.

As we increase $\alpha$ from zero we couple gravity, since $\alpha \sim G$.
With increasing $\alpha$ a first branch of compact boson star solutions 
evolves, where the scalar field
and the metric functions change continuously.
This first branch ends at a finite maximal value of $\alpha$,
$\alpha_{\rm max}$.
This is seen in Fig.~1d, where we exhibit
the value of the scalar function at the origin, $h(0)$
versus $\alpha$.
At $\alpha_{\rm max}$ a second branch bends backwards
towards smaller values of $\alpha$. While $h(0)$
now increases monotonically, $\alpha$ starts to exhibit
damped oscillations, tending towards a limiting value,
$\alpha_{\rm lim}$.

The scalar field and metric functions
of this family of solutions 
are exhibited in Figs.~1a-1c for a representative
set of values of $\alpha$, 
as marked by the dots in Fig.~1d.
In the oscillating regime of $\alpha$
the functions start to show a distinct behaviour,
deviating from the smooth first branch pattern.
In particular, the function $h(x)$ increases steeply
at the center of the boson star,
while the function $A(x)$ at the same time
decreases strongly at the center, tending towards zero
in the limit $\alpha \to \alpha_{\rm lim}$.

\begin{figure}[h!]
\begin{center}
\vspace{-0.5cm}
\mbox{\hspace{-0.5cm}
\subfigure[][]{\hspace{-1.0cm}
\includegraphics[height=.25\textheight, angle =0]{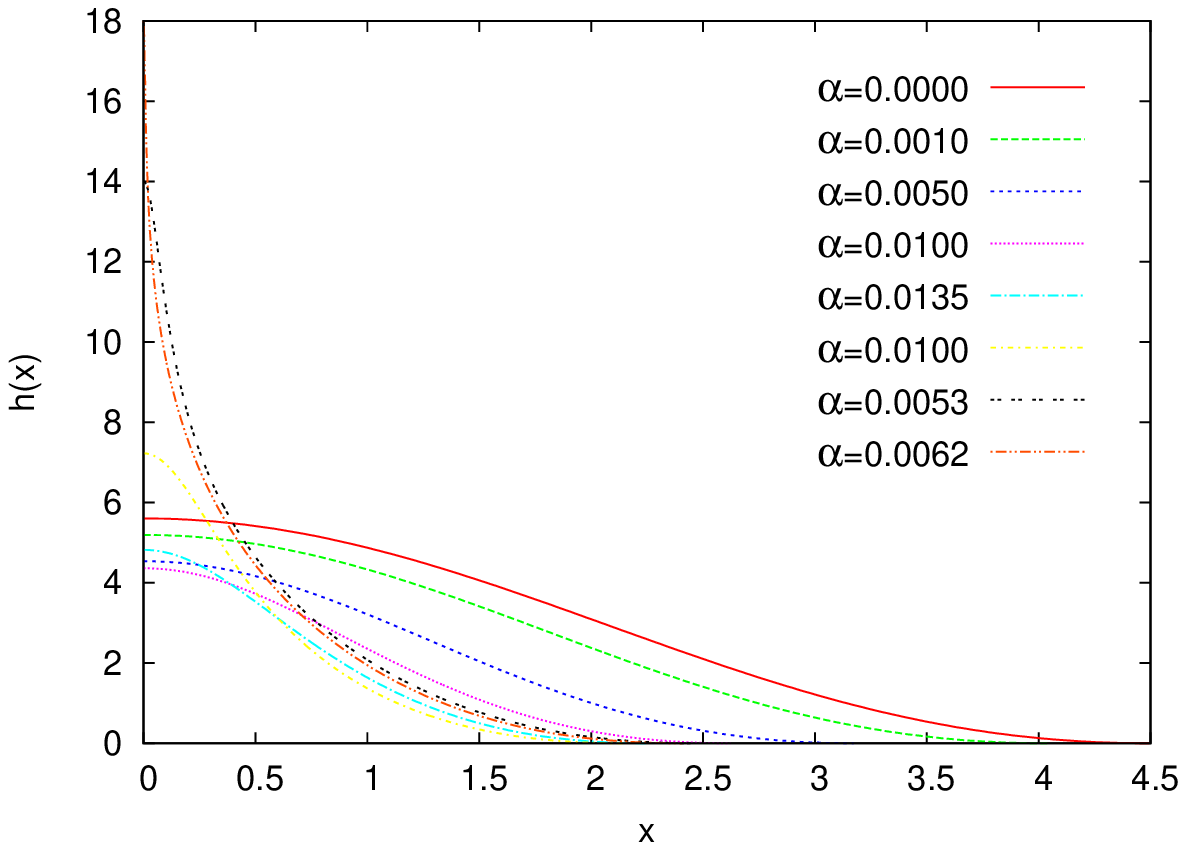}
\label{fig1a}
}
\subfigure[][]{\hspace{-0.5cm}
\includegraphics[height=.25\textheight, angle =0]{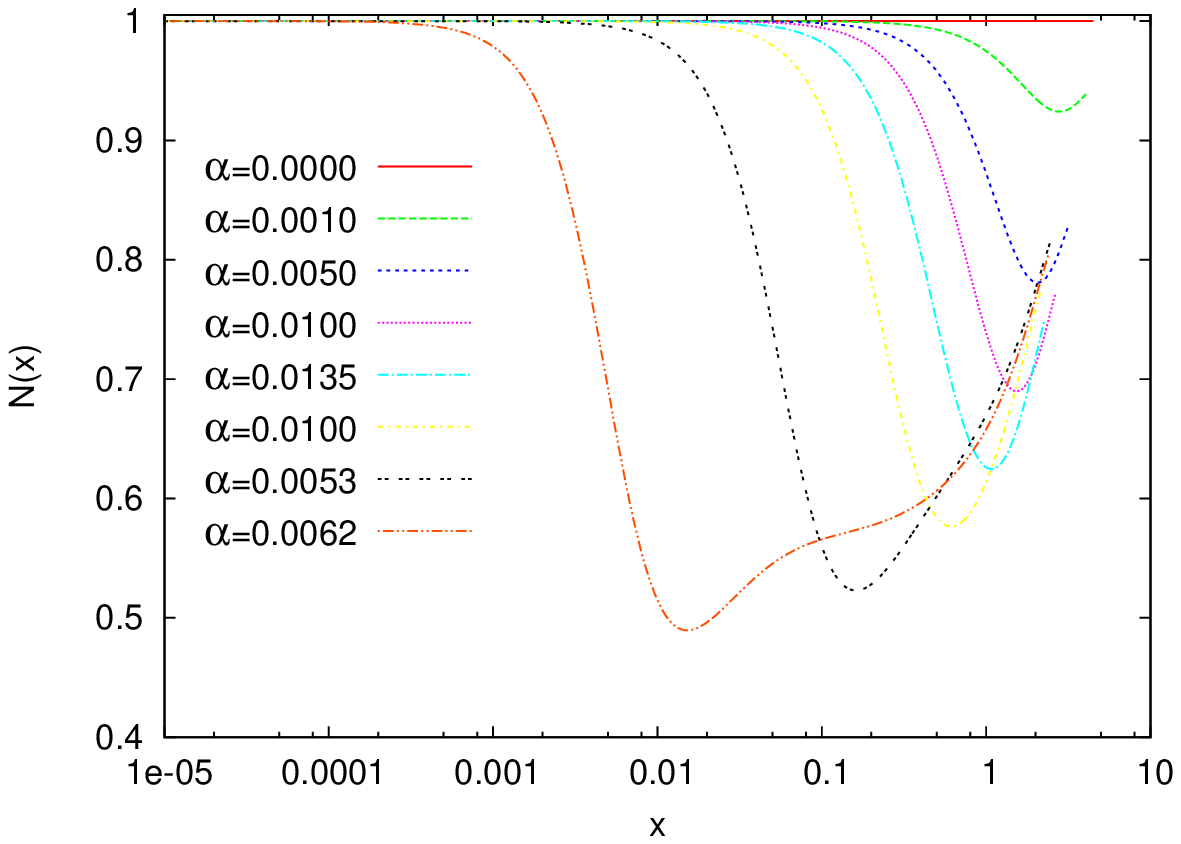}
\label{fig1b}
}
}
\mbox{\hspace{-0.5cm}
\subfigure[][]{\hspace{-1.0cm}
\includegraphics[height=.25\textheight, angle =0]{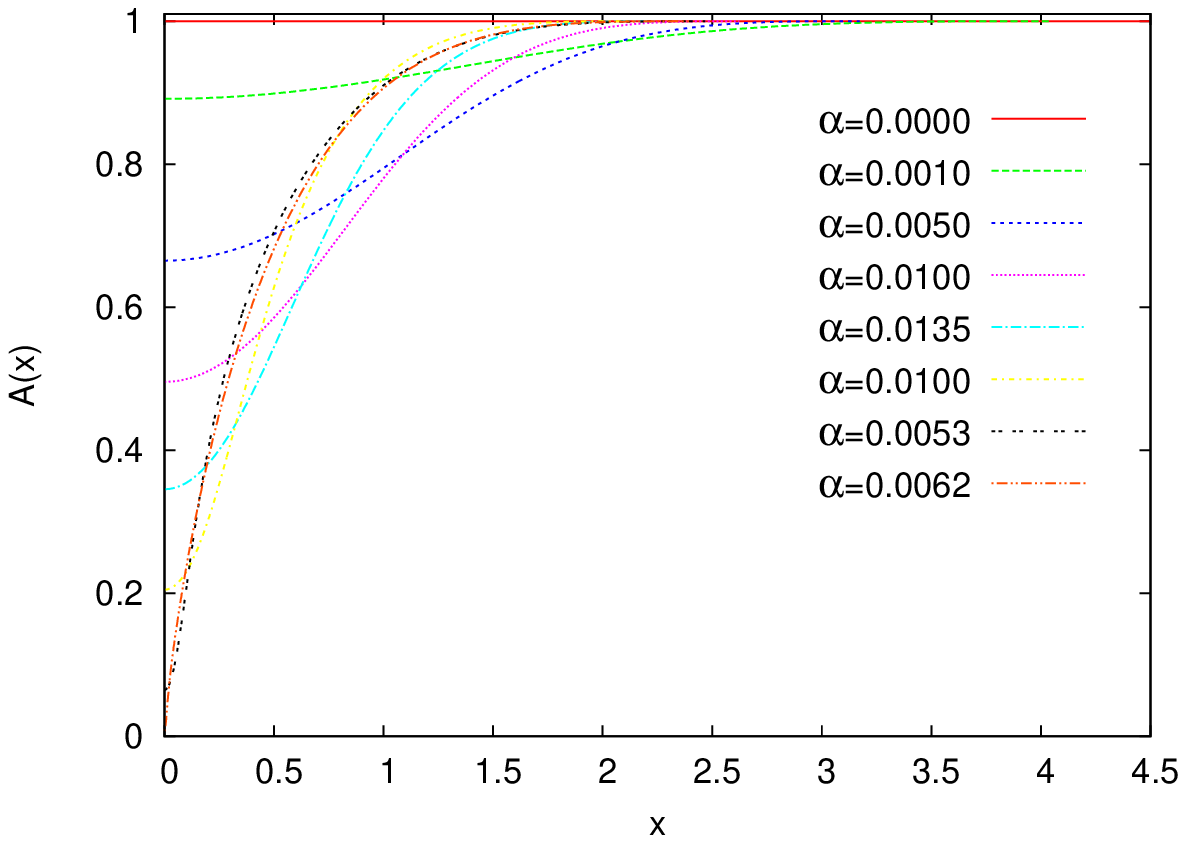}
\label{fig1c}
}
\subfigure[][]{\hspace{-0.5cm}
\includegraphics[height=.25\textheight, angle =0]{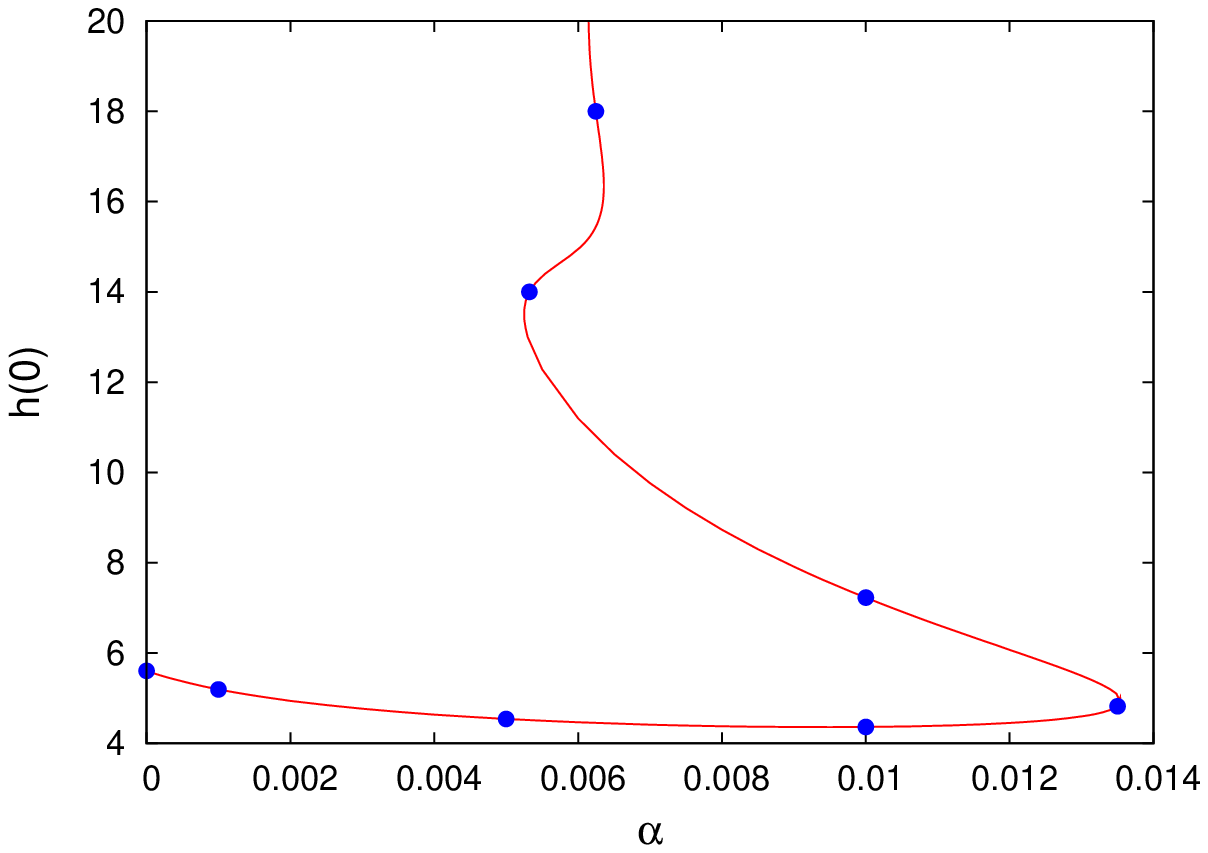}
\label{fig1d}
}
}
\end{center}
\vspace{-0.5cm}
\caption{
The scalar field function $h(x)$ (a),
the metric function $N(x)$ (b), and
the metric function $A(x)$ (c) versus $x$
for several values of $\alpha$.
Also shown are the values of the scalar field function
at the origin, $h(0)$, versus $\alpha$ (d). Here the dots
mark the set of values of $\alpha$ selected in (a)-(c).
\label{fig1}
}
\end{figure}

Let us next consider the relation between the parameter $\alpha$
and the boson star radius parameter $x_0$.
We exhibit $\alpha$ versus $x_0$ in Fig.2a.
The first branch starts from the flat space-time limit,
$\alpha=0$, where $\alpha$ increases with decreasing $x_0$.
The previously seen oscillating behaviour of $\alpha$
for the higher branches now translates into a spiralling
behaviour, when considered versus the radius parameter $x_0$.

\begin{figure}[h!]
\begin{center}
\vspace{-0.5cm}
\mbox{\hspace{-0.5cm}
\subfigure[][]{\hspace{-1.0cm}
\includegraphics[height=.25\textheight, angle =0]{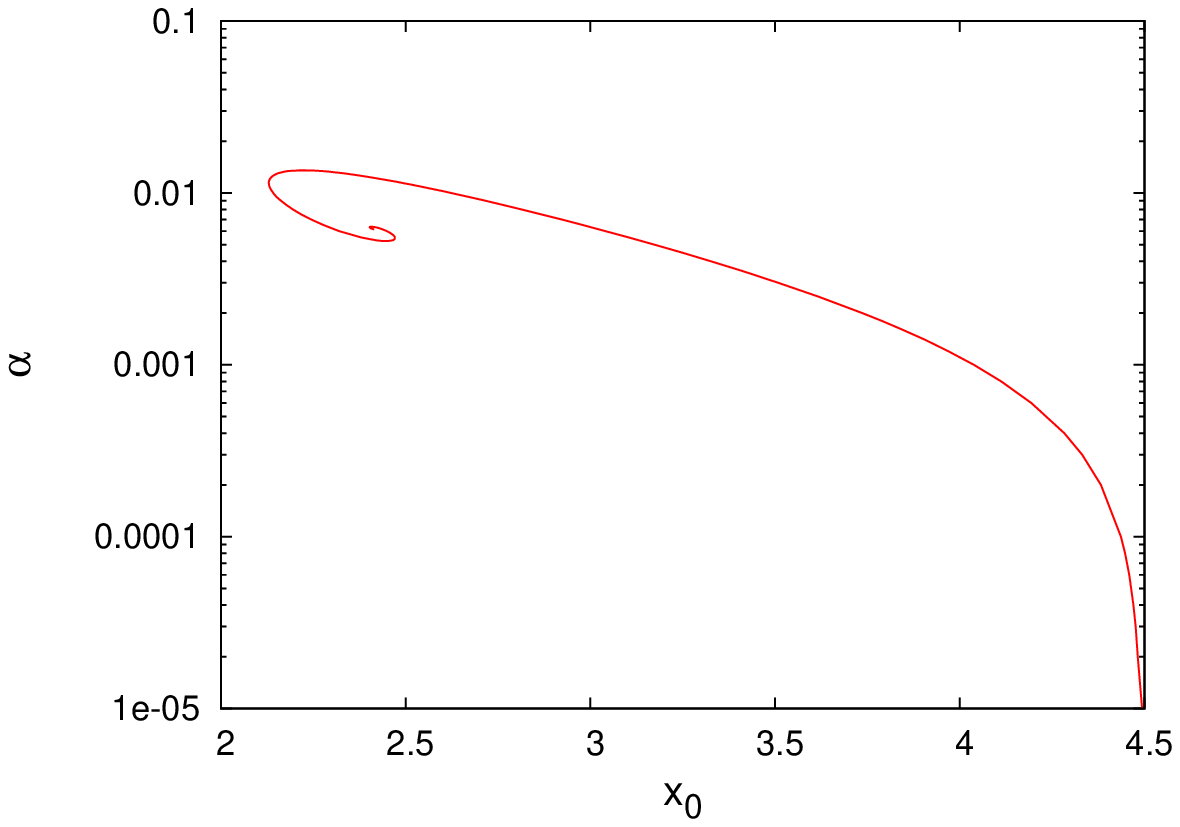}
\label{fig2a}
}
\subfigure[][]{\hspace{-0.5cm}
\includegraphics[height=.25\textheight, angle =0]{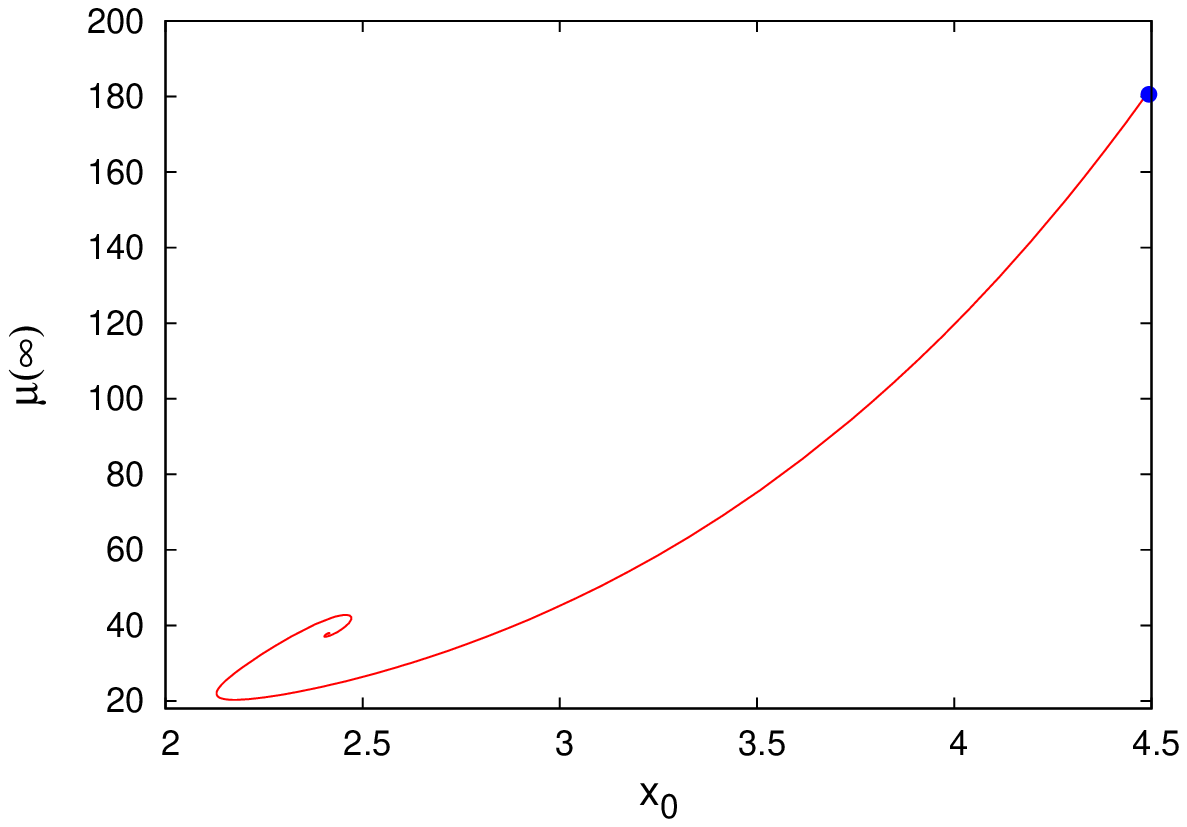}
\label{fig2b}
}
}
\mbox{\hspace{-0.5cm}
\subfigure[][]{\hspace{-1.0cm}
\includegraphics[height=.25\textheight, angle =0]{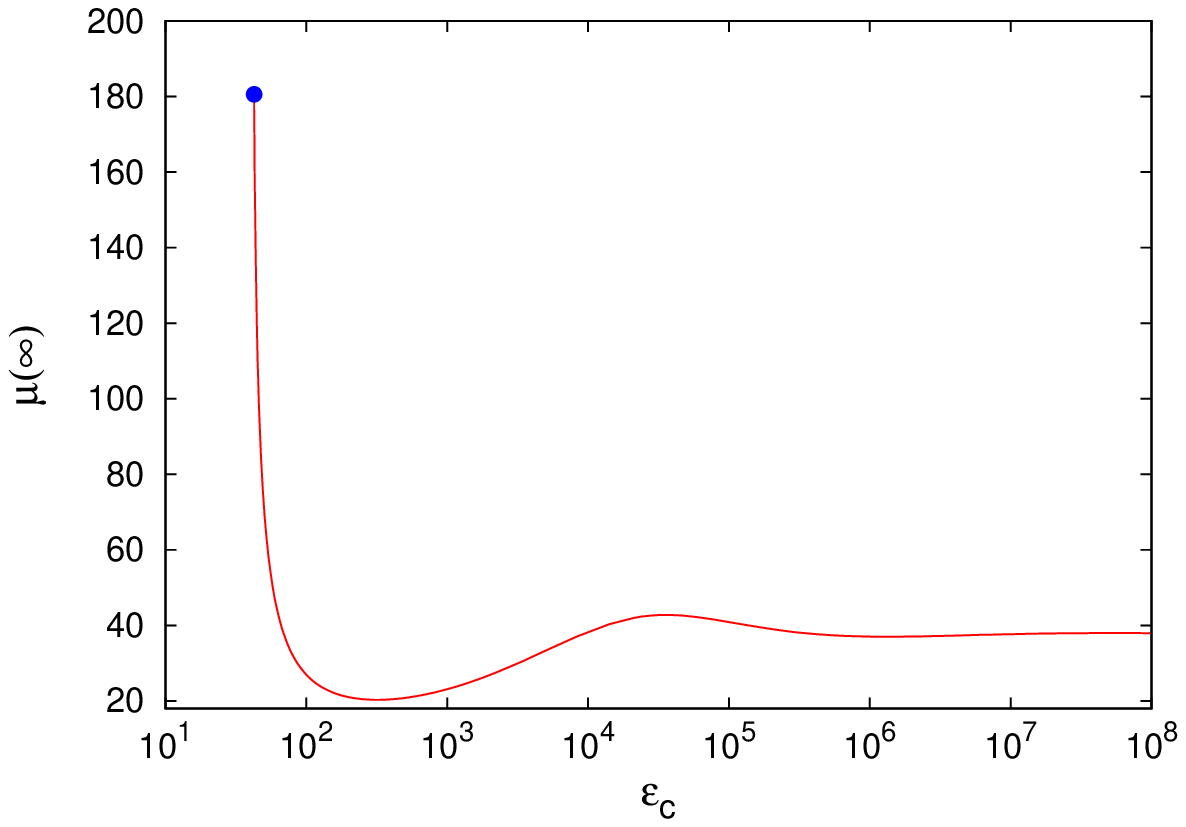}
\label{fig2c}
}
\subfigure[][]{\hspace{-0.5cm}
\includegraphics[height=.25\textheight, angle =0]{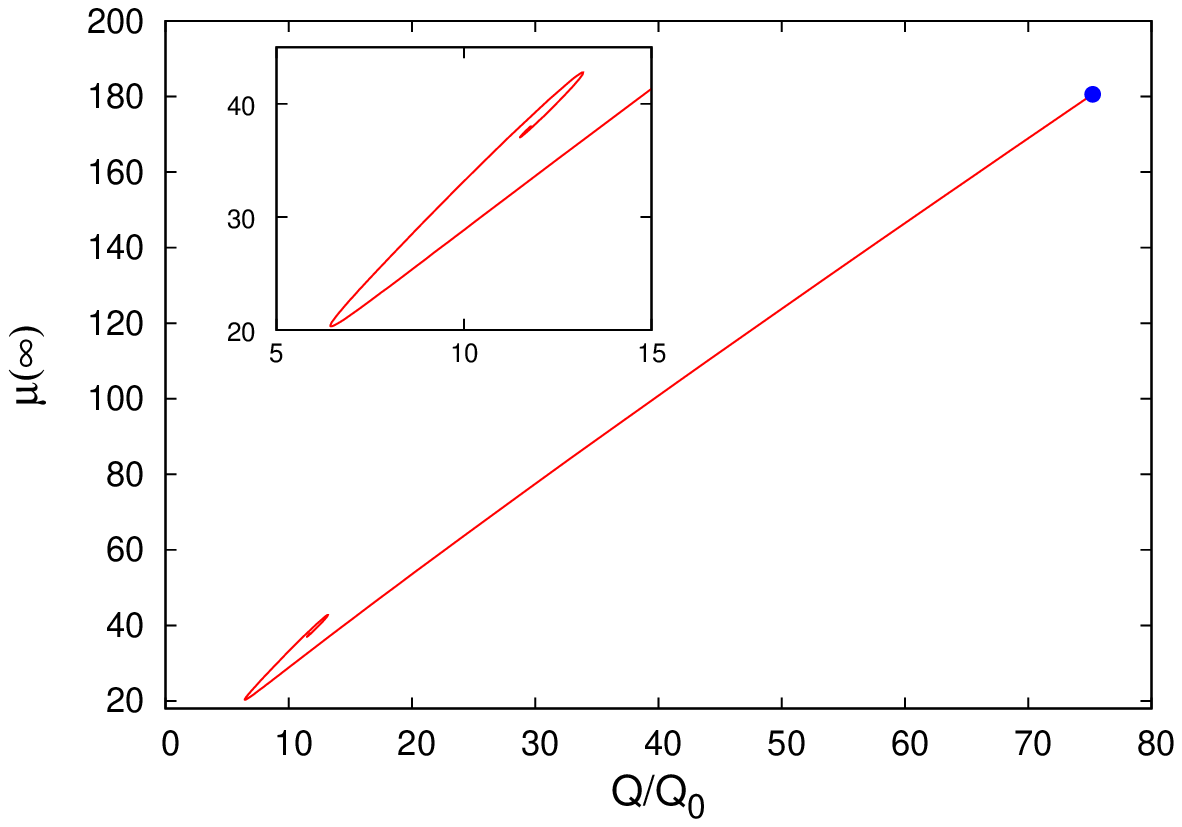}
\label{fig2d}
}
}
\end{center}
\vspace{-0.5cm}
\caption{
Coupling $\alpha$ versus radius parameter $x_0$ (a);
mass parameter $\mu$ versus radius parameter $x_0$ (b),
central density $\epsilon_c$ (c) and scaled particle number
$Q/Q_0$ (d). The dots mark the flat limit.
\label{fig2}
}
\end{figure}

Likewise, the mass parameter $\mu(\infty)$,
presented in Fig.2b, exhibits an analogous spiralling behaviour,
except that along the first branch the mass parameter
is decreasing with decreasing $x_0$,
and the spiral is inverted.
Clearly, a limiting value $\mu_{\rm lim}$ of the mass parameter is approached
at the center of the spiral, where $\alpha \to \alpha_{\rm lim}$.
The maximal value of $\mu$ corresponds to the flat space-time limit,
marked by a dot in the figure.

In Fig.2c we exhibit the mass parameter $\mu(\infty)$
versus the dimensionless central density $\epsilon_c$.
Here the mass parameter exhibits damped oscillations,
as it approaches $\mu_{\rm lim}$.
Finally, in Fig.2d, we consider the dependence of the
mass parameter on the scaled particle number.
Along the first branch, starting from the flat space-time limit,
the relation between the two quantities is almost linear.
Then again a spiralling behaviour sets in.

These figures with the dimensionless quantities
represent the full domain of existence
of the compact boson star solutions.
All physical solutions with dimensionful quantities
can be obtained from these dimensionless solutions by appropriate scaling.

To make contact with real stars and astrophysics, 
let us now discuss such sets of physical solutions.
To this end, we now consider the solutions for fixed models,
i.e., for fixed values of the coupling constant $\lambda$.
Moreover, we take the physical value for Newton's constant.
Then the value of $\alpha$ yields the value of the frequency $\omega$.
Together with $\mu(\infty)$ we then obtain the corresponding
value of the mass $M$ of the respective compact boson stars.
Likewise, by employing the radius parameter $x_0$
together with the frequency $\omega$
we obtain the physical value for the areal radius $R$
of the respective compact boson stars.

We exhibit the mass $M$ in units of the solar mass $M_\odot$
versus the radius $R$ in kilometers in Fig.~3a
for several values of the coupling constant $\lambda$,
determining the strength of the potential,
i.e., for several fixed models.
We observe, that the maximal value of the mass
and the maximal value of the radius increase
with decreasing coupling strength $\lambda$ of the potential.

Interestingly, the compact boson stars also correspond to 
physically compact stars, since for a mass on the order of the
solar mass their radius is on the order of ten(s) of kilometers,
thus corresponding in mass and size to neutron stars.
However, their mass increases with increasing size,
except in a small region close to the maximal value of the radius.
This behaviour is in contrast to neutron stars, where we observe
an increase of the mass with decreasing radius.
Instead, the observed dependence
of the mass on the radius is more in line with quark stars
\cite{Itoh:1970uw,Witten:1984rs,Farhi:1984qu,Alcock:1986hz}.
But such a behaviour has also been seen in boson stars
of the soliton type \cite{Kleihaus:2011sx}.

\begin{figure}[h!]
\begin{center}
\mbox{\hspace{-0.5cm}
\subfigure[][]{\hspace{-1.0cm}
\includegraphics[height=.25\textheight, angle =0]{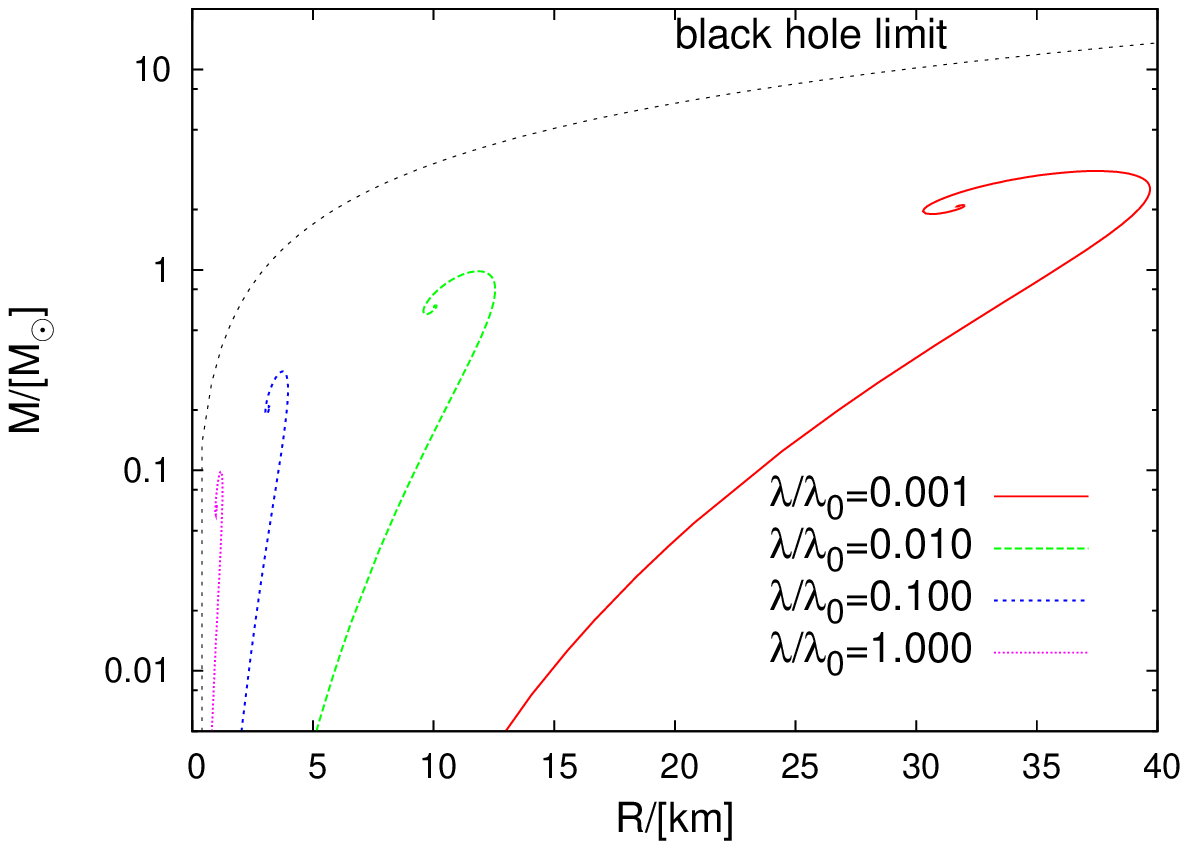}
\label{fig3a}
}
\subfigure[][]{\hspace{-0.5cm}
\includegraphics[height=.25\textheight, angle =0]{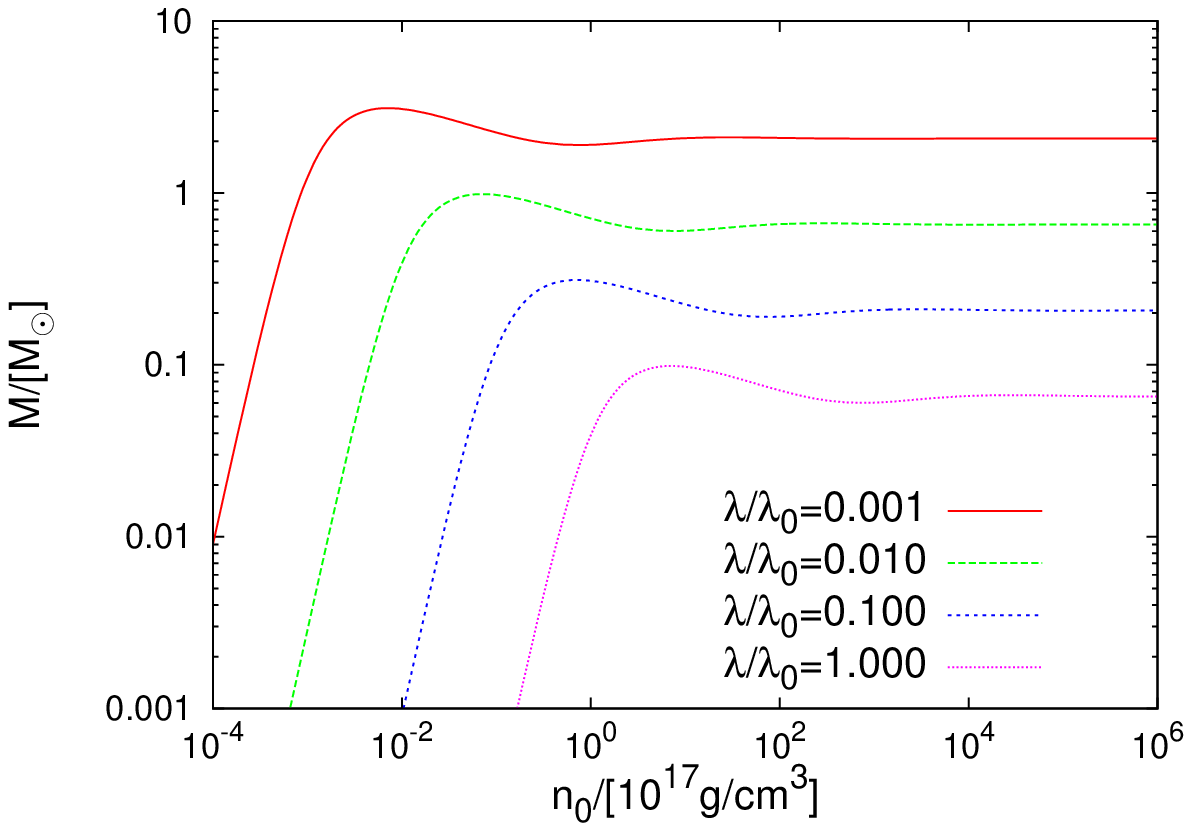}
\label{fig3b}
}
}
\end{center}
\vspace{-0.5cm}
\caption{
The mass in units of the solar mass
versus the areal radius (a) and the central mass density (b)
for several values of the potential strength $\lambda$
($\lambda_0 = \sqrt{ 2 \times 10^{30} {\rm kg}\, {\rm m}^{-3}\, {\rm s}^{-2} }$).
\label{fig3}
}
\end{figure}

As seen in Fig.~3a,
after reaching the maximal value of the mass
the solutions show the well-known spiralling behaviour
of highly compact stars.
For neutron stars
this behaviour is encountered, when the mass and size
of the stars approach the values of the
Schwarzschild black hole solution.
Therefore, for comparison, we have also
included the Schwarzschild limit in the figure.
While we are in the vicinity of this limit,
when the spirals are encountered for the compact boson stars,
we do not get very close.
In contrast, the soliton type boson stars
practically reach the black hole limit
\cite{Kleihaus:2011sx}.

It is also interesting to consider the dependence of the mass $M$
on the central mass density $n_0$ of the boson stars.
Therefore we exhibit this dependence for the same set of solutions
in Fig.~3b.
First, the mass reaches monotonically its absolute maximum,
and then exhibits the well-known oscillating behaviour,
also observed in neutron stars.
However, we note that the higher the mass at the maximum,
the lower the central mass density of the compact
boson stars.
We attribute this behaviour to the necessity of
lowering the potential strength in order to reach higher masses.

Let us finally address the stability of the configurations.
A mode analysis, as performed in the flat space limit
\cite{Lis:2009au}
is very involved. Therefore we will use arguments
from catastrophe theory to discuss the stability
\cite{Kusmartsev:2008py,Kusmartsev:1992,Tamaki:2010zz,Tamaki:2011zza,Kleihaus:2011sx}.
We will base our discussion on Fig.~3b, where the mass
is shown versus the central mass density for several values
of the potential strength.
In particular,
we choose the mass and the potential strength
as the two control parameters,
and we choose the central mass density as the single behavior variable
\cite{Tamaki:2010zz,Tamaki:2011zza,Kleihaus:2011sx}.

To analyze the stability of the compact boson stars, we start from the
\textit{equilibrium space} $\mathcal{M} = \{n_0, M, \lambda \}$,
as exhibited in Fig.~3b.
According to catastrophe theory,
the stability changes only at the turning points,
where ${\partial M}/{\partial n_0} = 0$
(for fixed values of $\lambda$). 
Thus passing a turning point means changing the stability
of the boson star configurations.
The first branch starts from a stable \cite{Lis:2009au} 
solution in flat space-time. 
Thus this branch should be stable, until the maximal value
of the mass is encountered.
At the maximum of the mass, however, stability should change,
and therefore the solutions in the spiral should be unstable.
In fact, we expect that a mode analysis will reveal,
that the solutions will become increasingly unstable inside the spiral,
collecting collecting more and more unstable modes.

\section{Conclusions and Outlook}

We have studied compact boson stars obtained from a V-shaped 
interaction potential.
Thus the scalar field is confined to a finite region.
By analyzing the scaled set of field equations,
we have obtained the full set of solutions,
consisting of two parts.
The first part corresponds to the branch of solutions
that emerges from the flat space-time solution,
and the second part,
depending on the physical quantities studied,
consists of a spiralling or oscillating set of solutions.

By specifying the constants of the model, we have translated
these dimensionless results into families of compact
boson star configurations whose masses and sizes
can be compared to real astrophysical objects.
Along the physical branch, emerging from the flat space-time solution,
the mass increases with the radius,
until a maximal value of the radius is reached.
This behaviour is unlike neutron stars and more akin to 
quark stars.
But as for these fermionic stars, the size of 
solar mass compact boson stars
in on the order of ten(s) of kilometers.

However, the masses of these
compact boson stars can be made arbitrarily high,
and their sizes increase accordingly, when the
potential strength is made sufficiently small.
These compact boson stars thus could represent
huge compact astrophysical objects,
not too far from their corresponding 
Schwarzschild black hole limit.

Employing arguments from catastrophe theory,
we have argued that the physical branch of these solutions,
that emerges from the stable flat-space solution,
is stable as well, while the remaining configurations are unstable.

As a next step we will include rotation
\cite{Arodz:2009ye}.
Rotating boson stars have been obtained before
for non-compact stars
\cite{Mielke:2000mh,Schunck:2003kk,Yoshida:1997qf,Kleihaus:2005me,Kleihaus:2007vk,Brihaye:2008cg}.
In particular, they exhibit a quantization relation
between the angular momentum and the particle number.
We expect, that
the inclusion of rotation in these compact boson stars
might lead to interesting new observations.

It should also be interesting
to construct interacting compact $Q$-balls
and boson stars, which should arise in the presence of several
complex scalar fields
\cite{Brihaye:2008cg,Brihaye:2007tn,Brihaye:2009yr}.

\vspace{0.5cm}
{\bf Acknowledgement}

\noindent
We would like to thank Meike List
and Eugen Radu for helpful discussions.
We gratefully acknowledge support by the DFG,
in particular, also within the DFG Research
Training Group 1620 ''Models of Gravity''.

\end{document}